\documentclass[prl,twocolumn,showpacs,amsmath,amssymb]{revtex4}
\usepackage[dvips]{graphicx}
\usepackage{latexsym}
\usepackage{amssymb}
\usepackage{amsmath}
\usepackage{amsbsy}

\begin{document}
\title{Kondo-Enhanced Andreev Tunneling in InAs Nanowire Quantum Dots}
\author{T. Sand-Jespersen, J. Paaske, B. M. Andersen, K. Grove-Rasmussen, H. I. J{\o}rgensen,
M. Aagesen, C. S{\o}rensen, P. E. Lindelof, K. Flensberg, J.
Nyg{\aa}rd.} \affiliation{Nano-Science Center, Niels Bohr Institute,
University of Copenhagen, Universitetsparken 5, DK-2100 Copenhagen,
Denmark}
\date{\today}
\begin{abstract}
We report measurements of the nonlinear conductance of InAs nanowire
quantum dots coupled to superconducting leads. We observe a clear
alternation between odd and even occupation of the dot, with
sub-gap-peaks at $|V_{sd}|=\Delta/e$ markedly stronger(weaker) than
the quasiparticle tunneling peaks at $|V_{sd}|=2\Delta/e$ for
odd(even) occupation. We attribute the enhanced $\Delta$-peak to an
interplay between Kondo-correlations and Andreev tunneling in dots
with an odd number of spins, and substantiate this interpretation by
a poor man's scaling analysis.
\end{abstract}
\pacs{72.15.Qm,73.21.La,73.23.-b,74.50.+r}

\maketitle

Since the discovery of the Kondo effect in quantum dots
(QD)\cite{GoldhaberNature:1998} this phenomenon has received
extensive theoretical and experimental attention\cite{a}. The effect
emerges for QD's coupled strongly to the leads when the total spin of
the electrons on the QD is non-zero, e.g. if it hosts an odd number
of electrons $N$. At temperatures below the so-called
Kondo-temperature, $T_K$, the conduction electrons in the leads
screen the spin through multiple co-tunneling spin-flip processes
resulting in a correlated many-body state which is experimentally
observable as an increased linear conduction through the dot. If the
leads to the QD consist of (s-wave) superconductors (S) the
conduction electrons form spin-singlet Cooper-pairs incapable of
flipping the dot spin and therefore the Kondo effect and
superconductivity constitute competing many-body effects.
\newline\indent
Recent developments in techniques for fabrication of quantum dot
systems have made it possible to produce S-QD-S
systems~\cite{Buitelaar:2002,Doh:2005,Vandam:2006,Jorgensen:2006,JarilloHerrero:2006}
enabling experimental studies of this intriguing interplay. In carbon
nanotubes~\cite{Buitelaar:2002}, it was found that the Kondo-state
persists when the energy needed for breaking the Cooper-pairs is
compensated by the energy gained in forming the Kondo-state
($T_{K}>\Delta$). Here $\Delta$ is the gap of the superconducting
leads. Otherwise the Kondo state is suppressed and the Kondo-induced
increase in the linear conductance disappears.
\newline\indent
Due to the superconductor gap, $\Delta$, in the leads, the nonlinear
conductance generally displays a quasi-particle tunneling peak at
bias-voltages $|V_{sd}| = 2\Delta/e$. At smaller voltages, transport
occurs through Andreev reflection (AR) processes where electrons
impinging on a superconducting electrode are reflected as holes upon
injecting a Cooper pair\cite{Andreev:1964}. This basic mechanism
gives rise to a rich subharmonic gap structure (SGS) in electrical
transport measurements which has been studied extensively in
superconducting weak links\cite{SGSweaklinks}, break
junctions~\cite{Sheer:1997}, and for quantum dots outside the Kondo
regime\cite{Buitelaar:2003}. Nevertheless, no experimental study of
the influence of Kondo correlations on the SGS has yet been reported
and this is the focus of the present work. We find that even when the
Kondo-peak in the linear conductance is suppressed by the
superconducting gap, a pronounced Kondo-enhancement of the leading
sub-gap peak in $dI/dV_{sd}$ emerges at $|V_{sd}|=\Delta/e$. We study
the characteristics of this peak and attribute it to a
Kondo-enhancement of the Andreev tunneling amplitude in dots with odd
occupancy.
\newline\indent
\begin{figure}
        \centering
        \includegraphics[width=8.5cm]{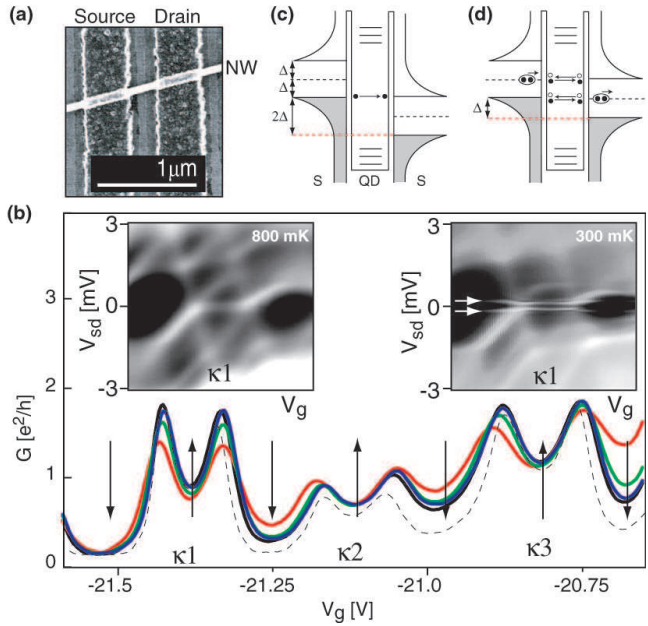}
        \caption{(Color online) (a) Scanning electron micrograph
        of a typical nanowire device. (b) Linear conductance in the Kondo-regime for temperatures
        $T=750\,\mathrm{mK (black)} - T=950\,\mathrm{mK (red)}$ and at $300\, \mathrm{mK}$ (dashed).
        The qualitative temperature dependence of the valley
        conductances for temperatures above $T_c$ are indicated by arrows.
        The leftmost inset shows $dI/dV_{sd}$ vs.
        $V_{sd}$ and $V_g$ for the region of $\kappa1$ at $800 \,
        \mathrm{mK} (> T_c)$ clearly showing the zero-bias Kondo ridge. At
        $300 \, \mathrm{mK}$, i.e., below $T_c$ (rightmost inset) the
        Kondo ridge is suppressed. Instead, a double peak
        structure is observed as indicated by the arrows (see text).
        (c),(d) Schematic illustration of the processes which leads
        to peaks at $|V_{sd}| = 2\Delta/e$ (direct quasiparticle tunneling, (c))
        and at $|V_{sd}| = \Delta/e$ (processes involving one Andreev
        reflection, (d)).}
        \label{FIG:SGS1}
\end{figure}%
Our devices are based on semiconducting InAs nanowires grown by
molecular beam epitaxy\cite{Aagesen:2007}. We have recently reported
on the Kondo effect in nanowire devices with normal metal Ti/Au
contacts\cite{Jespersen:2006} and the devices investigated in the
present study are identical with the exception that a superconducting
Ti/Al/Ti trilayer $(10/60/10 \mathrm{nm})$ is used for
contacting\cite{Doh:2005}. The superconducting transition temperature
$T_c \approx 750 \, \mathrm{mK}$ and critical magnetic field $B_c
\approx 250 \, \mathrm{mT}$ of the contact trilayer are determined
experimentally. Figure \ref{FIG:SGS1}(a) shows a scanning electron
micrograph of a device. The wire has a diameter $d\approx 70 \,
\mathrm{nm}$ and the electrode separation is $L\approx 300 \,
\mathrm{nm}$. As in Ref.\ \onlinecite{Jespersen:2006} the conductance
of the wire can be controlled by applying a voltage $V_g$ to the
degenerately doped Si substrate which acts as a back gate for
modulating the carrier concentration and barrier transparency. Below,
we focus on the intermediate coupling regime appropriate for Kondo
physics. The two-terminal conductance of the device is measured as a
function of applied source-drain bias and gate potential using
standard lock-in techniques.
\newline\indent
We first characterize the device with the contacts in the normal
state. Because of the high critical field of the Ti/Al/Ti contacts
and the very large $g$-factor of InAs\cite{Bjork:2005,Jespersen:2006}
$g \approx 9$, driving the contacts normal with a magnetic field will
significantly perturb the Kondo state\cite{a}. Therefore, we study
instead the device characteristics for temperatures above $T_c$.
Figure \ref{FIG:SGS1}(b) shows the linear conductance $G$ as a
function of $V_g$ for temperatures $750\,\mathrm{mK} - 950\,
\mathrm{mK}$ when the contacts are normal (solid lines). A series of
overlapping Coulomb peaks are observed and the temperature dependence
of the valley conductances are indicated by the arrows. In four
valleys the conductance decreases upon lowering the temperature, as
expected for Coulomb blockade. For the three valleys labeled $\kappa
1 - \kappa 3$ the opposite behavior is observed, signifying Kondo
physics. The left inset is a grey scale plot of the differential
conductance $dI/dV_{sd}$ vs.\ $V_g$ and $V_{sd}$ (stability diagram)
measured at $800 \, \mathrm{mK} (> T_c)$. It shows the familiar
pattern of Coulomb diamonds (charging energy $E_C \approx 1.5 \,
\mathrm{meV}$, level spacing $\Delta E \sim 1 \, \mathrm{meV}$) and
confirms the presence of a high conductance Kondo ridge around zero
bias through the diamond $\kappa 1$. The black dashed line shows $G$
vs.\ $V_g$ measured at $300\, \mathrm{mK} (<T_c)$. Instead of
continuing their increase as expected for the Kondo effect (without
superconductivity) the valley conductances $G_v$ of $\kappa 1 -
\kappa 3$ decrease below their values at $950 \, \mathrm{mK}$. This
result is consistent with the findings of Ref.\ \cite{Buitelaar:2002}
and shows that the binding energy $\sim k_{B}T_{K}$ of the Kondo
states $\kappa 1- \kappa 3$ is lower than the binding energy of the
superconducting Cooper pairs $\Delta \approx 1.75k_BT_c$. Here $k_B$
is Boltzmann's constant and the Kondo temperature $T_K$ is therefore
smaller than $\Delta/k_{B}\approx 1.3 \, \mathrm{K}$ in all three
charge-states $\kappa 1-\kappa 3$\cite{comment1}.
\newline\indent
The suppression of the Kondo state is confirmed by the disappearance
of the Kondo ridge in the stability diagram measured at $300 \,
\mathrm{mK}$ shown for $\kappa 1$ in the rightmost inset to Fig.\
\ref{FIG:SGS1}(b).
\begin{figure}[b]
        \centering
        \includegraphics[width=8.5cm]{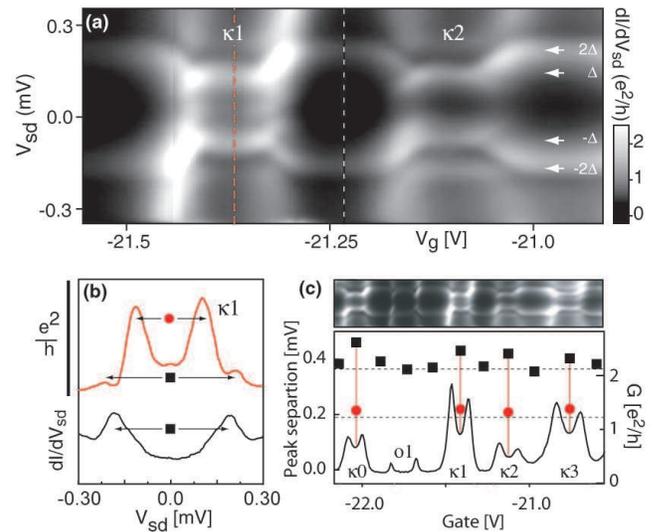}
        \caption{(Color online) (a) Stability diagram for the $V_g$-region
        of $\kappa1$ and $\kappa2$ and $|V_{sd}| \le 0.35 \,
        \mathrm{mV}$ exhibiting a pronounced even-odd periodicity in the sub-gap
        structure. The approximate positions of the dominant peaks at
        $V_{sd} = \pm \Delta/e, \pm 2\Delta/e$ are indicated.
        (b) $dI/dV_{sd}$ vs.\ $V_{sd}$ through the middle
        of two neighboring diamonds (lines in (a)) showing the enhanced peaks at $|V_{sd}| \approx
        \Delta/e$ for the Kondo-diamond (curves offset for clarity). (c) Bias spectroscopy for a larger
        $V_g$-region showing the enhanced $\Delta$-peak for four Kondo diamonds. Main panel shows
        the linear conductance (rightmost axis) and the distance between
        $\pm \Delta$ and $\pm 2\Delta$ sub-gap peaks in the middle of the diamonds
         (leftmost axis, symbols as in (b)). Horizontal dashed lines
        indicate the position of $2\Delta$ and $4\Delta$.
        }
        \label{FIG:SGS2}
\end{figure}%
The finite bias peaks indicated by the arrows are observed throughout
the stability diagram and appear symmetrically around $V_{sd} = 0\,
\mathrm V$ (see also Fig.\ \ref{FIG:SGS2}). These are manifestations
of the superconducting state of the contacts where peaks are expected
at $|V_{sd}| = 2\Delta/e$ when the density of states at the gap edges
line up as illustrated in Fig.\ \ref{FIG:SGS1}(c). The transport for
energies below the superconducting energy gap $(|V_{sd}| <
2\Delta/e)$ is mediated by multiple Andreev reflections (MAR) and
peaks in $dI/dV_{sd}$ are expected each time a new Andreev process
becomes accessible\cite{SGSweaklinks}. The stability diagram in Fig.\
\ref{FIG:SGS2}(a) shows a detailed measurement of these low-bias
features for the Coulomb diamonds of $\kappa 1$ and $\kappa 2$. Close
to the degeneracy points of the Coulomb diamonds a complicated peak
structure is observed, since in this region the MAR occur resonantly
through the gate-voltage dependent dot
level\cite{Buitelaar:2003,LevyYeyati:1997,Johansson:1999}.
\newline\indent
We restrict our discussion to the middle region of the Coulomb
diamonds where the peak positions are largely gate-independent and
transport occurs by co-tunneling between the two superconductors. In
this case, MAR peaks are expected at $|V_{sd}| = 2\Delta/ne, n =
1,2,\dots$ with intensities determined mainly by the effective
transparency of the device. The process responsible for the peaks at
$|V_{sd}| = \Delta/e$ ($n=2$) involves one Andreev reflection and is
shown schematically in Fig. \ref{FIG:SGS1}(d). The black(lower) trace
in Fig.\ \ref{FIG:SGS2}(b) shows the differential conductance along
the white dashed line in Fig.\ \ref{FIG:SGS2}(a) through the middle
of an even-$N$ diamond. As expected, peaks are observed at $|V_{sd}|
= 2\Delta/e$ with fainter shoulders at $|V_{sd}| = \Delta/e$.
However, as seen in Fig.\ \ref{FIG:SGS2}(a) the SGS in the odd-$N$
diamonds of the suppressed Kondo ridges $\kappa 1$ and $\kappa 2$ is
clearly modified. Unexpectedly, the peaks at $|V_{sd}| = \Delta/e$
are more than 5 times larger than the peaks at $|V_{sd}| = 2\Delta/e$
as emphasized by the red(upper) trace in Fig.\ \ref{FIG:SGS2}(b)
which shows a trace through the middle of the $\kappa 1$-diamond.
This contrasts the expectations for simple tunneling between the two
superconductors, and the presence of the Kondo effect in the normal
state points towards electron-electron correlations as the origin of
the modified SGS\cite{comment2}.
\newline\indent
Further support of the connection between the Kondo effect and the
enhanced $\Delta$-peak is provided in Fig.\ \ref{FIG:SGS2}(c): The
lower panel shows the linear conductance (right axis) at
$V_{sd}=0\,\mathrm V$ over a $V_g$-range of 5 odd-$N$ and 6 even-$N$
Coulomb valleys. In the three odd-$N$ valleys $\kappa 1 - \kappa 3$
from Fig.\ \ref{FIG:SGS1}(b) and an additional one, $\kappa 0$, the
Kondo effect was observed in the normal state, and as seen in the
upper panel, the enhanced $\Delta$-peak is observed in the SGS of
each Coulomb diamond. In the remaining diamonds, including $o1$ with
odd-$N$ (which did not show the Kondo effect in the normal state),
the conventional SGS is observed. Thus, the effect is connected to
Kondo correlations rather than to the number of electrons on the dot.
In the lower panel of Fig.\ \ref{FIG:SGS2}(c) the separations between
the $-2\Delta$ and $+2\Delta$-peaks (squares), and between the
$-\Delta$ and $+\Delta$-peaks (circles) peaks are extracted for each
Coulomb valley (left axis). The separations of the $\pm
2\Delta$-peaks depends slightly on $N$ and are increased in the
$\kappa$-valleys with respect to the even-$N$ values, however, the
$\Delta$-peaks in the $\kappa$-valleys always appear at exactly half
the separation of the corresponding $2\Delta$-peaks.
\begin{figure}
        \centering
        \includegraphics[width=8.5cm]{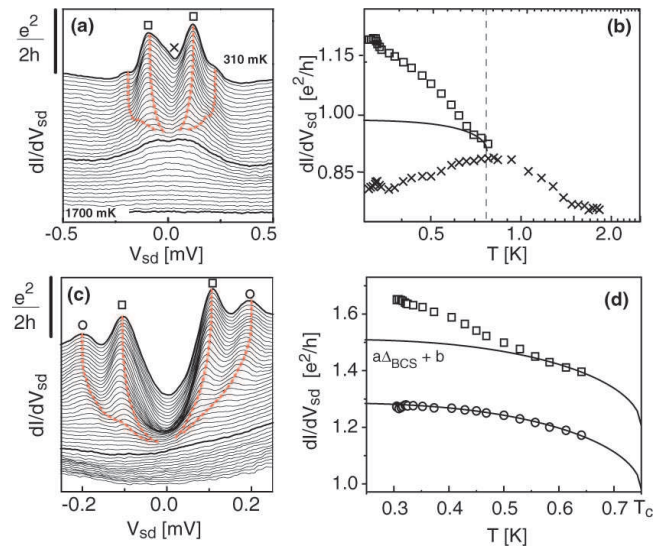}
        \caption{(Color online) (a), $dI/dV_{sd}$ vs.\ $V_{sd}$ through $\kappa 1$ for different
temperatures $310 \,\mathrm{mK}$ - $1700 \,\mathrm{mK}$ (offset for
clarity). Bold trace shows the Kondo peak at $T = 800 \,\mathrm{mK}$
and the formation of the $\Delta/e$-peak can be followed. Symbols
correspond to traces in (b) showing the suppression of the valley
conductance (crosses) and the continuing increase of the average
heights of the $\Delta$ peaks below $T_c$ (squares). Solid line shows
a fit to the gap function $\Delta_{BCS}(T)$ expected in the absence
of correlations. (c) Measurements as in (a) from another device. The
$\Delta$ and $2\Delta$ peaks can be clearly resolved and the average
peak heights are extracted in (d). The evolution of the $2\Delta$
peaks (circles) agree with $\Delta_{BCS}(T)$ (solid line), however,
as in (b) the continued increase of the $\Delta$-peak is not captured
(squares, offset for clarity).}
        \label{FIG:SGS3}
\end{figure}%
\newline\indent
One of the most distinct features of the conventional Kondo effect is
provided by the temperature dependence of the Kondo ridge, and to
investigate further the origin of the $\Delta$-peak we have studied
the temperature dependence of the SGS. Figure \ref{FIG:SGS3}(a) shows
$dI/dV_{sd}$ vs.\ $V_{sd}$ through the middle of $\kappa1$ for
temperatures $1700\, \mathrm{mK} - 310 \, \mathrm{mK}$. Upon lowering
the temperature, the initial formation of the Kondo peak is observed
for $T > T_c$, followed by the formation of the sub-gap peaks at
$|V_{sd}| = \Delta/e$ with shoulders at $|V_{sd}| = 2\Delta/e$ for
$T<T_c$. The peak positions follow the expected temperature
dependence of the BCS gap $\Delta_{BCS}(T)$ (indicated by red lines).
To allow for a better comparison of the evolution of the $\Delta$ and
$2\Delta$ peaks we include in panel (c) a similar measurement
performed for a suppressed Kondo ridge in a different device where
the four peaks can be clearly distinguished. In panels (b) and (d)
the temperature dependencies of the average peak heights and the
valley conductances have been extracted. In both cases the
$\Delta$-peaks increase with roughly constant slope, whereas the
$2\Delta$-peaks of the measurement in panel (c) saturate below $\sim
0.5 \, \mathrm{K}$. In the absence of interactions the temperature
dependence is governed by the gap, and we expect the measured peak
heights to be proportional to $\Delta_{BCS}(T)$\cite{Wolf:1985}. The
solid lines in (b),(d) show such fits and $\Delta_{BCS}(T)$ indeed
describes the behavior of the $2\Delta$-peak in panel (d). For the
enhanced $\Delta$-peaks, however, the continuing strengthening for
decreasing temperatures is not captured, thereby adding further
evidence to the importance of correlations for the origin of these
peaks.
\newline\indent
In the middle of the Coulomb diamonds, charge-fluctuations are
strongly suppressed and electrons traverse the dot via cotunneling
processes with a tunneling amplitude of the order of $J_{LR}\sim
t_{L}^{\ast}t_{R}/E_{C}$, where $t_{L,R}$ are the tunneling
amplitudes from the dot to the two leads. This means that the even
occupied dot can be viewed as an effective superconducting
single-mode junction with transparency
$\alpha\sim(\nu_{F}J_{LR})^{2}\ll 1$, which has been studied in
Ref.~\onlinecite{SGStheory} ($\nu_F$ denotes the density of states at
the Fermi-levels). In single-mode junctions with low transparency,
the quasiparticle tunneling conductance peak at $|V_{sd}|=2\Delta/e$
dominates over the sub-gap peaks, as observed in our even Coulomb
diamonds. In the case of odd occupation, however, transport occurs
via {\it exchange}-(co)tunneling which gives rise to a Kondo-enhanced
transparency of the order of $1/\ln^{2}(\Delta/k_{B}T_{K})$. Since
the leading sub-gap peak at $V_{sd}=\Delta/e$ can exceed the
$2\Delta$-peak for moderate values of the
transparency~\cite{SGStheory}, our observations are consistent with a
Kondo-enhanced transparency of the junction. This argument is valid
only for $\Delta\gg k_{B}T_{K}$, but clearly the spin-full dot holds
the premise for a Kondo-enhanced transparency, which provides a
simple understanding of the $\Delta$-peak dominating the
$2\Delta$-peak in the odd diamonds which supported a Kondo resonance
for temperatures above $T_C$.
\newline\indent
The logarithmic enhancement of the transparency can be established
from a poor man's scaling analysis~\cite{Anderson:1970} of the Kondo
model with superconducting leads and $\Delta\gg k_{B}T_{K}$. The
exchange coupling, $\nu_{F}J_{LR}$, grows stronger as the conduction
electron bandwidth, $D$, is reduced, and terminates at
$1/\ln(\Delta/k_{B}T_{K})$ when $D$ reaches $\Delta$. Interestingly,
the scaling also generates a new anomalous operator independent of
the impurity spin. The effective low-energy (time-dependent)
Hamiltonian thus contains an Andreev tunneling term,
$A_T[\,e^{2i(\mu_{L}\!-\!\mu_{R})t}{\textstyle\sum_{k',k}}
c^{\dagger}_{Lk'\uparrow}c^{\dagger}_{Lk\downarrow}\!+(L\leftrightarrow
R)]+h.c.$, where $\mu_{L/R}$ denotes the chemical potentials and the
explicit time-dependence from the applied bias has been gauged into a
phase-factor on the electron-operators. The current is readily
calculated to second order in $A_{T}$ and sets in with a step at
$V=\Delta$. Schematically, the coupled scaling equations for $J$ and
$A_T$ take the form
\begin{align}
\frac{dA_T}{d\ln D}&=
-\frac{3}{4}\frac{\Delta}{D}J^2-4\frac{\Delta}{D}A_T^2,\nonumber\\
\frac{dJ}{d\ln D}&= -J^2+2\frac{\Delta}{D}JA_T,\nonumber
\end{align}
leaving out all lead-indices $(L,R)$ and step-functions determining
the energy-scales beyond which the various terms nolonger contribute
to the flow. For $D\gg\Delta$, only $J$ grows logarithmically whereas
the flow of $A_T$ is prohibited by extra factors of $\Delta/D$
deriving from the virtual propagation of a Cooper-pair near the band
edge. For $D\ll\Delta$, however, these factors of $\Delta/D$ would
rather enhance the flow and lead to a divergence of $A_{T}$. We
postpone the full analysis of these scaling equations to a future
publication~\cite{FuturePub:2007}, and merely note here that the
alignment of Fermi-level with gap-edge for $|V_{sd}|=\Delta/e$ might
permit the flow to continue to strong coupling, which in turn may
lead to the anomalous temperature dependence which we observe for the
$\Delta$-peak height. Further analysis of the finite-bias scaling
equations will clarify this issue. Finally, we note that the
Andreev-tunneling operator is also generated in the even diamonds by
simple potential scattering, but in this case the $J^2$-term driving
the enhancement of the transparency is missing.
\newline\indent
In summary, we have discovered a pronounced alternation of the
strength of the leading sub-gap conductance peak between even and odd
occupied quantum dots coupled to superconducting leads. We have
observed this effect in $15$ suppressed Kondo ridges in two different
devices. We ascribe the enhancement of the $\Delta$-peaks for odd
occupations to a Kondo-enhanced Andreev tunneling amplitude.
Furthermore, we have found that, unlike the $2\Delta$-peak, the
$\Delta$-peak height does not saturate with $\Delta(T)$ when lowering
the temperature.
\newline\indent
{\it Note:} After completing this work, we have become aware of an
independent, parallel study of the above phenomenon in a different
materials system, carbon nanotubes, by A. Eichler {\it et
al.}~\cite{Eichler:2007}.
\newline\indent
{\it Acknowledgement} This work was supported by the EC FP6 funding
(Project FP6-IST-003673) and the Danish Agency for Science,
Technology and Innovation.


\vspace*{-4mm}
\begin{thebibliography}{99}
\vspace*{-6mm}
\bibitem{GoldhaberNature:1998} D. Goldhaber-Gordon {\it et al.},
Nature {\bf 391}, 156 (1998).
%
\bibitem{a} L. P Kouwenhoven and L. I. Glazman, Phys. World {\bf 14}, 33 (2001).
%
\bibitem{Buitelaar:2002} M. R. Buitelaar, T. Nussbaumer, and C. Sch{\"o}nenberger, Phys. Rev. Lett. {\bf 89}, 256801 (2002).
%
\bibitem{Doh:2005} Y. J. Doh {\em et al.}, Science {\bf 309}, 272 (2005).
%
\bibitem{Vandam:2006} J. A. van Dam {\em et al.}, Nature, {\bf 442}, 667 (2006).
%
\bibitem{Jorgensen:2006} H. I. J\o rgensen {\em et al.}, Phys. Rev. Lett. {\bf 96}, 207003 (2006).
%
\bibitem{JarilloHerrero:2006} P. Jarillo-Herrero, J. A. van Dam, and L. P. Kouwenhoven, Nature {\bf 439}, 953 (2006).
%
\bibitem{Andreev:1964} N. F. Andreev, Sov. Phys. JETP {\bf 19}, 1228 (1964).
%
\bibitem{SGSweaklinks} M. Octavio, M. Tinkham, G. E. Blonder, and T. M. Klapwijk, Phys. Rev. B {\bf
27}, 6739 (1983); K. Flensberg, J. B. Hansen, and M. Octavio, Phys.
Rev. B 38, 8707-8711 (1988).
%
\bibitem{Sheer:1997} E. Scheer {\em et al.}, Phys. Rev. Lett. {\bf 78}, 3535 (1997).
%
\bibitem{Buitelaar:2003} M. R. Buitelaar {\em et al.}, Phys. Rev. Lett. {\bf 91}, 057005 (2003).
%
\bibitem{Aagesen:2007} M. Aagesen, {\em et al.} in preparation.
%
\bibitem{Jespersen:2006} T. S. Jespersen {\em et al.}, Phys. Rev. B {\bf 74}, 233304 (2006).
%
\bibitem{Bjork:2005} M. T. Bj{\"o}rk {\em et al.}, Phys. Rev. B {\bf 72}, 201307(R) (2005).
%
\bibitem{comment1} In agreement with Ref.~\cite{Buitelaar:2002} we have also observed Kondo ridges
that survive the transition to superconducting contacts and are
further enhanced by the superconductor as expected for $T_K >
\Delta/k_B$.
%
\bibitem{LevyYeyati:1997} A. L. Yeyati, J. C. Cuevas, A. L\'{o}pez-D\'{a}valos, and A. Martin-Rodero, Phys. Rev. B {\bf 55}, R6137 (1997).
%
\bibitem{Johansson:1999} G. Johansson, E. N. Bratus, V. S. Shumeiko, and G. Wendin, Phys. Rev. B {\bf 60}, 1382 (1999).
%
\bibitem{comment2} We note, that a similar enhanced
$\Delta/e$-peak in the SGS of a carbon nanotube QD is evident in
Refs.~\onlinecite{Buitelaar:2002,Buitelaar:2003} and in K.
Grove-Rasmussen, H. I J{\o}rgensen, P.E. Lindelof, cond-mat/0601371;
Proc. Int. Symp. on Mesoscopic Superconductivity and Spintronics
2006, (World Scientific Publishing, 2007), where possible connection
to Kondo effect was suggested but no conclusions drawn.
%
\bibitem{SGStheory}D. Averin and A. Bardas, Phys. Rev. Lett. {\bf 75}, 1831
(1995); E. N. Bratus, V. S. Shumeiko, and G. Wendin, Phys. Rev. Lett.
{\bf 74}, 2110 (1995); J. C. Cuevas, A. Martin-Rodero, and A. L.
Yeyati, Phys. Rev. B {\bf 54}, 7366 (1996).
%
\bibitem{Wolf:1985} E. L. Wolf, Principles of Electron Tunneling Spectroscopy, (Oxford University Press, 1985), p.\ 113.
%
\bibitem{Anderson:1970}P. W. Anderson, J. Phys. C {\bf 3}, 2436 (1966).
%
\bibitem{FuturePub:2007} J. Paaske, B. M. Andersen and K. Flensberg (unpublished).
%
\bibitem{Eichler:2007}A. Eichler, {\em et al.}, cond-mat/0703082.
\end{thebibliography}
\end{document}